\pdfoutput=1
\documentclass[conference]{IEEEtran}
\usepackage{amsfonts}
\usepackage{graphicx}
\usepackage{color}
\usepackage{amsmath,amsfonts,amssymb,amsthm,epsfig,epstopdf,url,array}
\usepackage{url,textcomp}
\usepackage{cite}
\usepackage[misc]{ifsym}
\usepackage{verbatim}
\usepackage{subfigure}
\usepackage{caption}
\usepackage{algorithm}
\usepackage{algorithmicx}
\makeatletter
\newcommand\multiline[1]{\parbox[t]{\dimexpr\linewidth-\ALG@thistlm}{#1}}
\makeatother
\usepackage{algpseudocode}
\def\BibTeX{{\rm B\kern-.05em{\sc i\kern-.025em b}\kern-.08em
    T\kern-.1667em\lower.7ex\hbox{E}\kern-.125emX}}

\begin{document}
\IEEEoverridecommandlockouts
\title{Graph Convolutional Network Enabled Power-Constrained HARQ Strategy for URLLC}

\author{
\IEEEauthorblockN{Yi~Chen\IEEEauthorrefmark{1},
        Zheng~Shi\textsuperscript{\Letter}\IEEEauthorrefmark{1},
        Hong~Wang\IEEEauthorrefmark{2},
        Yaru~Fu\IEEEauthorrefmark{3},
        Guanghua~Yang\IEEEauthorrefmark{1},
        Shaodan Ma\IEEEauthorrefmark{4},
       and Haichuan Ding\IEEEauthorrefmark{5}}\\
\IEEEauthorrefmark{1}School of Intelligent Systems Science and Engineering, Jinan University, Zhuhai 519070, China.\\
\IEEEauthorrefmark{2}School of Communication and Information Engineering, Nanjing University of Posts and Telecommunications, \\Nanjing 210003, China.\\
\IEEEauthorrefmark{3}School of Science and Technology, Hong Kong Metropolitan University, Hong Kong SAR, China .\\
\IEEEauthorrefmark{4}The State Key Laboratory of Internet of Things for Smart City, University of Macau, Macau.\\
\IEEEauthorrefmark{5}School of Cyberspace Science and Technology, Beijing Institute of Technology, Beijing 100081, China.
\thanks {This work was supported in part by National Natural Science Foundation of China under Grants 62171200, 62171201, and 62261160650, in part by Guangdong Basic and Applied Basic Research Foundation under Grant 2023A1515010900, in part by Zhuhai Basic and Applied Basic Research Foundation under Grant ZH22017003210050PWC, in part by the Hong Kong Research Matching Grant (RMG) in the Central Pot under Project No. CP/2022/2.1, in part by the Major Talent Program of Guangdong Provincial under Grant 2019QN01S103, in part by the Science and Technology Development Fund, Macau SAR under Grants 0087/2022/AFJ and SKL-IOTSC(UM)-2021-2023, and in part by Open Research Foundation of National Mobile Communications Research Laboratory of Southeast University under Grant 2023D01. (\emph{Corresponding Author: Zheng Shi.})}
}

\maketitle
\begin{abstract}
In this paper, a power-constrained hybrid automatic repeat request (HARQ) transmission strategy is developed to support ultra-reliable low-latency communications (URLLC). In particular, we aim to minimize the delivery latency of HARQ schemes over time-correlated fading channels, meanwhile ensuring the high reliability and limited power consumption. To ease the optimization, the simple asymptotic outage expressions of HARQ schemes are adopted. Furthermore, by noticing the non-convexity of the latency minimization problem and the intricate connection between different HARQ rounds, the graph convolutional network (GCN) is invoked for the optimal power solution owing to its powerful ability of handling the graph data. The primal-dual learning method is then leveraged to train the GCN weights. Consequently, the numerical results are presented for verification together with the comparisons among three HARQ schemes in terms of the latency and the reliability, where the three HARQ schemes include Type-I HARQ, HARQ with chase combining (HARQ-CC), and HARQ with incremental redundancy (HARQ-IR). To recapitulate, it is revealed that HARQ-IR offers the lowest latency while guaranteeing the demanded reliability target under a stringent power constraint, albeit at the price of high coding complexity.

\end{abstract}
\begin{IEEEkeywords}
Graph neural networks, HARQ-IR, power allocation, time-correlated fading channels
\end{IEEEkeywords}

\IEEEpeerreviewmaketitle
\section{Introduction}\label{sec:intro}
\IEEEPARstart Nowadays, ultra-reliable low-latency communications (URLLC) have become an unprecedented paradigm shift to support the mission-critical internet-of-things (IoT) applications \cite{9826826}. For instance, as per the 3rd generation partnership project (3GPP), a 32 byte packet is expected to transmit within 1 ms along with a reliability of at least 99.999\%. To confront this stringent requirement, hybrid automatic repeat request (HARQ) is one of the key enabling technologies that provide reliable transmissions to combat channel fading. On the basis of different encoding and decoding techniques, HARQ can be divided into three types, namely Type-I HARQ, HARQ with chase combining (HARQ-CC), and HARQ with incremental redundancy (HARQ-IR). In essence, HARQ sacrifices the delay performance to improve reliability, which inevitably hinders its widespread applications in supporting URLLC. To overcome this shortcoming, HARQ should be properly designed with more flexibility to accommodate diverse requirements of latency and reliability.

The optimal design of HARQ schemes has been extensively studied in the literature. To name a few, in \cite{6697940}, the outage probability of HARQ-CC was minimized by imposing a constraint on the average power consumption. The asymptotic outage probability was used to enable the optimal power allocation with geometric programming (GP). The similar method was then applied to solve the minimization of the expected energy consumption for HARQ-IR given a maximum allowable outage tolerance in \cite{9473556}. Moreover, in \cite{8362650}, the goodput of HARQ-IR was maximized through the joint optimization of the transmission powers and transmission rate under an average power constraint. The joint optimization of powers and rate was further considered to maximize the energy efficiency of HARQ schemes, which were solved in closed-form with the Karush-Kuhn-Tucker (KKT) conditions. Furthermore, the optimization of various HARQ-assisted systems has also received considerable research interest lately. To be specific, the power efficient design was considered for HARQ-CC aided non-orthgonal multiple access (NOMA) systems in \cite{9089002}, wherein successive convex approximation (SCA) was used to provide the optimal power solution. HARQ-NOMA-assisted short packet communications were investigated in \cite{9211722}, where a genetic algorithm was applied to optimize the power levels in power-constrained and reliability-constrained scenarios. 
Apart from high reliability and limited power consumption, the guarantee of low latency is also of profound significance to realize URLLC, while this topic was rarely examined except in a few existing works. Particularly,
in \cite{10015032}, the age of information (AoI) was minimized for HARQ-IR-assisted multi-RIS systems while ensuring power and outage constraints. 
In addition, the authors in \cite{9606298} maximized the overall rates of enhanced mobile broadband (eMBB) users in HARQ-assisted grant free systems under the constraint of a maximum tolerable probability of delay bound violation. However, independent fading channels were generally assumed in these works \cite{9473556, 9089002, 9211722, 10015032, 9606298} whose results are inapplicable to the correlated fading channels. Due to the frequent occurrence of time-selective fading channels, there is an urgent need to propose a proper latency assurance strategy for HARQ schemes over correlated fading channels. 

In order to minimize the delivery latency while guaranteeing the high reliability, this paper proposes a power-constrained HARQ strategy by considering time-correlated fading channels. The simple asymptotic outage expressions of HARQ schemes are adopted to avoid heavy computational burden. Unfortunately, the optimization problem still cannot be easily solved by  using the off-the-shelf tools due to the fractional objective function and non-convex constraints. This has inspired us to explore the use of artificial intelligence (AI) techniques. By taking into account that the transmit power allocated in the current HARQ round is affected by the previous HARQ rounds regardless of the subsequent HARQ rounds, and the time correlation takes place among fading channels, it is natural to come up with graph neural networks to capture this special transmission structure. The graph convolutional network (GCN) is then invoked for the optimal power allocation of HARQ by treating different HARQ rounds and channel correlation as graph nodes and edges, respectively. The trainable GCN weights are updated by using primal-dual learning approach. Finally, the latency and the reliability performance of the GCN-enabled power allocation strategy for three HARQ schemes was investigated through numerical experiments, where the three HARQ schemes include Type-I HARQ, HARQ-CC, and HARQ-IR. Last but not least, despite its high coding complexity, HARQ-IR can provide the lowest latency while ensuring reliable performance within strict power constraints.


The rest of the paper is structured as follows. Section \ref{sec:model} elaborates on the system model and formulates the latency minimization problem for power-constrained HARQ schemes. In Section \ref{sec:gcnmodel}, a GCN-enabled power allocation strategy is designed to solve the optimization problem. Section \ref{sec:EXPERIMENTS} verifies the effectiveness of the proposed strategy through numerical experiments. Finally, concluding remarks are drawn in Section \ref{sec:CONC}.
\section{System Model}\label{sec:model}
This paper considers a point-to-point HARQ-aided URLLC system. To begin, the system model is delineated in this section, including HARQ schemes, signal transmission model, and problem formulation.
\subsection{HARQ Schemes}
HARQ can be classified into three categories according to different coding operations, including Type-I HARQ, HARQ-CC, and HARQ-IR. More specifically, for both Type-I HARQ and HARQ-CC, the same codeword is delivered in all HARQ rounds. At the receiver side, Type-I HARQ decodes its message by solely relying on the currently received codeword, while HARQ-CC combines the erroneously received codewords for maximal-ratio combining (MRC). Undoubtedly, HARQ-CC outperforms Type-I HARQ due to the fact that even the failed packets contain useful information. Unlike Type-I HARQ and HARQ-CC, HARQ-IR transmits codewords with different redundancy among all HARQ rounds. Hence, a long codeword is first chopped into several sub-codewords at the transmitter, which will be sent one by one upon retransmission request. At the receiver, the previously received codewords are then concatenated to form a long codeword for the joint decoding. Owing to the high encoding/decoding complexity, HARQ-IR achieves the superior performance of reliability.


\begin{figure}[htbp]
    \centering
    \includegraphics[width=8cm]{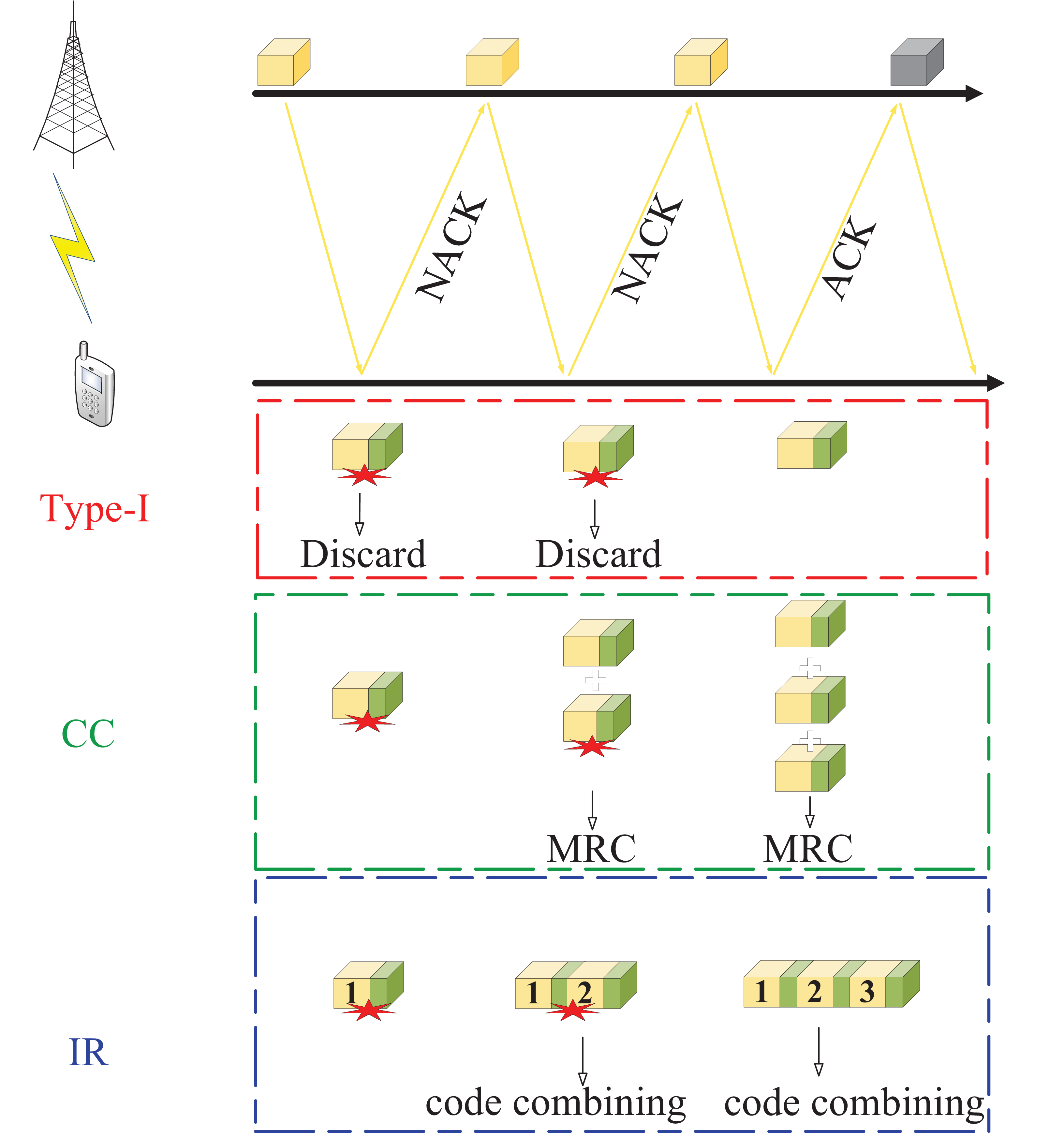}
    \caption{An example of HARQ transmissions.}
    \label{FIG1} 
\end{figure}

\subsection{Signal Transmission Model}
By considering blocking fading channels, the received signal in the  $k$-th HARQ round can be expressed as
\begin{equation}\label{eqn:receive signal}
{{\bf{y}}_k} = {h_k}{{\bf{x}}_k} + {{\bf{n}}_k},
\end{equation}
where ${{\bf{x}}_k}$ denotes the $k$-th codeword of length $M$ with average power $p_k$, and ${{\bf{n}}_k}$ denotes a complex additive white Gaussian noise vector with zero mean vector and identity covariance matrix, i.e., ${{\bf{n}}_k} \sim {\cal CN}({\bf 0},{{\bf{I}}_{ M}})$ 
,
${h_k}$ refers to the channel coefficient of the $k$-th transmission. To avoid large transmission latency under unfavorable fading channels, the maximum number of HARQ rounds for sending each message is limited up to $K$. Due to the frequent occurrence of the correlation among fading channels \cite{5692978}, the time-correlated Rayleigh fading channels are used to model $h_k$ as
\begin{equation}\label{eqn:channel model}
{h_k} = {\xi    _k}\left(\sqrt {1 - {\rho ^{2(k + \delta  - 1)}}} {\alpha _k} + {\rho ^{k + \delta  - 1}}{\alpha _0}\right),
\end{equation}
where the factor $\rho$ measures the intensity of the time correlation between channel coefficients, $\delta$ and $ {\xi _k}^2$ denote the feedback delay and the average power of the channel, respectively, ${\alpha _0},{\alpha _1},...,{\alpha _k}$ are mutually independent and obey complex normal distributions with zero mean and unit variance.

According to \eqref{eqn:receive signal}, the received signal-to-noise ratio (SNR) in the $k$-th transmission can be obtained as
\begin{equation}\label{eqn:SNR}
{\gamma _k} = {p_k}{\left| {{h_k}} \right|^2}, \, k\in [1,K], 
\end{equation}
where ${p_k}$ denotes the average transmit power in the $k$-th HARQ round. 


\subsection{Problem Formulation of Power-Constrained HARQ for URLLC}
The mission-critical IoT applications usually emphasize stringent constraints of reliability and latency \cite{9810267}. Besides, the IoT devices  are frequently equipped with non-rechargeable battery which cannot provide continuous power supply. The energy consumption of IoT networks mainly comes from the radio frequency communication module. Hence, the transmit powers should be optimally devised to prolong the lifetime of IoT networks. In this paper, we aim at the accommodation of low latency as well as high reliability via the power allocation in different HARQ rounds. To proceed, we denote by $N_b$ and $B$ the total number of information bits and the bandwidth, respectively. The delivery latency of these information bits is thus calculated as
\begin{equation}\label{eqn:delay_def}
\tau = \frac{N_b}{\eta B},
\end{equation}
where $\eta$ denotes the effective spectral efficiency. The spectral efficiency of HARQ scheme can be estimated by using the long term average throughput (LTAT) of HARQ, which can be obtained as \cite{6879267}
\begin{equation}\label{eqn:throught2}
\eta  = \frac{{R(1 - {P_{out,K}})}}{{1 + \sum\limits_{k=1}^{K - 1} {{P_{out,k}}} }},
\end{equation}
where $R = b/M$ and $b$ denote the preset transmission rate  and the number of original information bits, respectively. In addition, $P_{out,k}$ refers to the outage probability after $k$ HARQ rounds. With the definition of the latency, the latency minimization problem of the power-constrained HARQ while guaranteeing its high reliability can be formulated as
\begin{equation}\label{eqn:min_delay}
\begin{array}{*{20}{cl}}
{\mathop {\min }\limits_{{p_1}, \cdots ,{p_K}} }&\tau \\
{\rm s.t.}&{{P_{out,K}} \le \varepsilon },\\
{}&{p_{\rm avg} \le \bar p,}
\end{array}
\end{equation}
where the reliability is ensured by imposing a constraint on the outage probability, and $\varepsilon$ denotes the maximum acceptable outage tolerance, ${\bar p}$ denotes the maximum allowable total transmit power, and $p_{\rm avg}$ denotes the average transmit power that is evaluated as
\begin{equation}
    p_{\rm avg} = \sum\limits_{k = 1}^K {{p_k}{P_{out,k - 1}}},
\end{equation}
and we stipulate $P_{out,0} = 1$. It should be mentioned that the power allocation is optimally designed by only utilizing the statistical channel state information (CSI) to avoid frequent signaling interactions and conserve time.

\section{GCN-Enabled Power Allocation Strategy}\label{sec:gcnmodel}
Although the exact outage expressions of HARQ systems were derived in \cite{7448651,7833558,7959548}, the outage expressions involve the summation of an infinite number of special functions. Such complex representations entail a high computational burden for the optimal design. In order to overcome this issue, the asymptotic expressions of the outage probabilities are leveraged in the optimal design. As derived in \cite{7448651,7833558,7959548}, the asymptotic outage probabilities of three types of HARQ schemes are given by 
\begin{equation}\label{eqn:pout_asy}
{P_{out,K}} = \left\{ {\begin{array}{*{20}{c}}
{{\varsigma _K}{{({2^R} - 1)}^K},}&{{\rm{Type-I}}}\\
{{\varsigma _K}{{({2^R} - 1)}^K}/K!,}&{{\rm{CC}}}\\
{{\varsigma _K}{\mathcal G_K}(R),}&{{\rm{IR}}}
\end{array}} \right.,
\end{equation}
where $\Gamma \left(  \cdot  \right)$ denotes the Gamma function, 
${\varsigma _K} = \frac{{{{\left( {\ell\left( {\rho ,K} \right)} \right)}^{ - 1}}}}{{\prod\nolimits_{k = 1}^K {{p_k}{\xi _k}^2} }}$, ${\ell}(\rho ,K)$ quantifies the effect of correlation that is given by
\begin{equation}\label{eqn:L_func}
\begin{array}{l}
{\ell}(\rho ,K) = \left(1 + \sum\limits_{k = 1}^K {\frac{{{\rho ^{2(k + \delta  - 1)}}}}{{1 - {\rho ^{2(k + \delta  - 1)}}}}} \right)\prod\limits_{k = 1}^K {\left(1 - {\rho ^{2(k + \delta  - 1)}}\right)},
\end{array}
\end{equation}
it should be noted that $\rho \ne 1$, and $\mathcal G_K(R)$ reads as
\begin{equation}\label{eqn:g_func}
\begin{array}{l}
{{\cal G}_K}(R) = {( - 1)^K} + {2^R}\sum\limits_{k = 0}^{K - 1} {{{( - 1)}^k}\frac{{{{(R\ln 2)}^{K - k - 1}}}}{{(K - k - 1)!}}}.
\end{array}
\end{equation}

Unfortunately, due to the correlation among different transmissions, the fractional form of the objective function, and non-convexity of constraints, the optimization problem in \eqref{eqn:min_delay} still cannot be easily solved by means of the classical optimization methodologies, such as CVX tools. The success of the application of GCNs in power allocation policy learning for wireless networks \cite{9252917} has motivated us to develop GCN-based power allocation for power-constrained HARQ schemes. It is clear that different HARQ rounds and correlations among them can be represented by graph nodes and edges, respectively. 
It is noteworthy that GCN is suitable herein due to its ability of exploiting the graph structure to process data. In what follows, a GCN-enabled power allocation scheme is detailed.
\subsection{GCN-Based Power Allocation}

To enable the GCN-based power allocation, each HARQ round is modeled as a graph node. Moreover, as aforementioned that the statistical CSI is used, the graph edges can be characterized by the correlation coefficients among fading channels. More specifically, the channel correlation coefficient matrix ${\bf H}$ is calculated as
\begin{equation}\label{eqn:H_def_co}
    {\bf H} = 
    \left[  {\alpha}_{ij} \right]_{1 \le i,j\le K},
\end{equation}
where the element ${\alpha _{ij}} $ is
\begin{equation}
    {\alpha _{ij}} = \left\{ {\begin{array}{*{20}{c}}
{{\rm{E}}\left\{ {{h_i}^{*}{h_j}} \right\},}&{i \le j}\\
{0,}&{\rm else}
\end{array}} \right.,
\end{equation}
wherein the superscript ${*}$ denotes the complex conjugate operation and $\rm E\{\cdot\}$ is the expectation operation. The rationale of ${\alpha _{ij}}=0$ for $i>j$ is due to the fact the $j$-th HARQ round cannot be influenced by the $i$-th HARQ round. With the time-correlated channel model in \eqref{eqn:channel model}, the diagonal entries of ${\bf H}$ can be calculated as ${\rm{E}}\left\{ {{{\left\| {{h_i}} \right\|}^2}} \right\} = {\xi _i}^2$ and its off-diagonal entries are given by ${\rm{E}}\left\{ {{h_i}^{*}{h_j}} \right\} = {\xi _i}{\xi _j}{\rho ^{i + j + 2\delta  - 2}}$ \cite{6094281}. 
Apparently, ${\bf H}$ can be treated as the adjacency matrix in the directed graph network.



\begin{figure}[htbp]
    \centering
    \includegraphics[width=8cm]{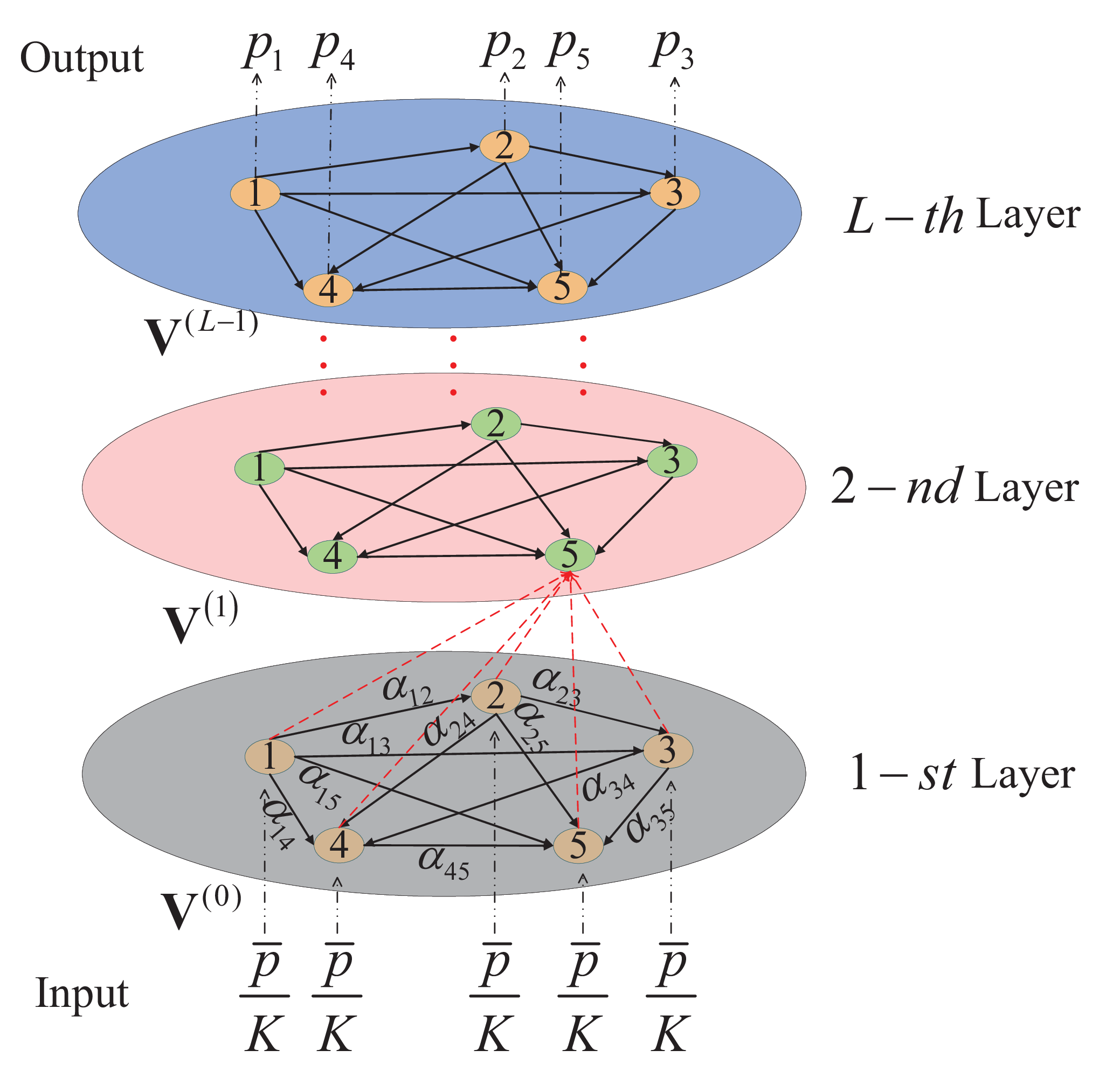}
    \caption{A general $L$-layer GCN structure for the power-constrained HARQ schemes with $K=5$.} 
    \label{FIG2} 
\end{figure}

For tractability, the power policy functional space ${\bf p} = ( {{p_1},{p_2},...,{p_K}} )$ is parameterized by using a graph neural networks. More specifically, the power allocation policy is defined as ${\bf p}({\bf H}) = \Psi({\bf H};{\bf W})$, where $\Psi$ represents a $L$-layer GCN with trainable weights ${\bf W}$. Instead of optimizing ${\bf p}$, the neural network parameters ${\bf W}$ need to be optimally determined through the primal-dual learning approach \cite{9072356}. As shown in Fig. \ref{FIG2}, a $L$-layer GCN structure for the power-constrained HARQ schemes with $K=5$ is given as an example. With the input ${\bf V}^{(0)} = \frac{\bar p}{K} {\bf 1}_K$ to $\Psi({\bf H};{\bf W})$, 
the $(l+1)$-th layer features  ${\bf V}^{(l + 1)}$ of GCN are updated by following the layer-wise propagation rule as
\begin{equation}\label{eqn:GNNfunc}
{{\bf{V}}^{(l + 1)}} = \sigma_l\left({{\bf{D}}^{ - \frac{1}{2}}}{\bf{H}}{{\bf{D}}^{ - \frac{1}{2}}}{{\bf{V}}^{(l)}}{{\bf{W}}^{(l)}}\right),
\end{equation}
where ${\bf 1}_K$ stands for an all-ones column vector, ${{\bf{D}}} = {\rm diag}({\bf{H}})$ denotes the degree matrix, ${{\bf{V}}^{(l)}} \in {\mathbb R^{K\times n_{l}}}$ is the matrix of node features in the $l$-th layer, ${{\bf W}^{(l)}} \in \mathbb R^{n_l\times n_{l+1}}$ is the trainable weight matrix in the $l$-th layer, $\sigma _l( \cdot )$ defines the activation function. Accordingly, the output of the $L$-th layer of the GCN is our desired power allocation policy. 
\subsection{Primal-Dual Learning Approach}
In order to train the neural network weights ${\bf W}$, the iterative primal-dual learning approach is applied in this paper. Moreover, by realizing that the maximum allowable outage probability is generally very low (e.g., $\varepsilon=10^{-2}$), we apply the logarithm transformation to the outage probability, i.e., $\log {P_{out,K}}$, in order to amplify the effect of the outage constraint and meanwhile accelerate learning and alleviate the over-fitting.
By taking this transformation into consideration, the Lagrangian of problem \eqref{eqn:min_delay} is formulated as
\begin{multline}\label{eqn:dual prrobelm}
\mathcal L_{\Psi}({\bf{W}},\lambda ,\upsilon ) = \tau + \lambda (\log {P_{out,K}} - \log \varepsilon )\\
 + {\upsilon }\left(\sum\limits_{k = 1}^K {{p_k}{P_{out,k-1}}} - \bar p \right),
\end{multline}
where $\lambda  \ge 0$ and $\upsilon  \ge 0$ are the Lagrangian multiplier associated with the two constraints in problem \eqref{eqn:min_delay}. It is noteworthy that $\tau $, ${P_{out,K}}$, and $\bf p$ are the functions of the trainable parameters $\bf W$. According to the gradient descent algorithm, the neural network parameters $\bf W$ at step $s$ can be updated as

\begin{equation}\label{eqn:optweight}
\begin{array}{l}
{{\bf{W}}_{s + 1}} = {{\bf{W}}_s} - {\theta _{{\bf{W}},s}} {\nabla _{\bf{W}}}{{\rm E}_{\rho  \sim \mathcal A}}\left\{\mathcal L({{\bf{W}}_s},{\lambda_s} ,{\upsilon_s})\right\},
\end{array}
\end{equation}
where the term ${\theta _{{\bf{W}},s}}$ denotes the step size at the $s$-th iteration and the actual time correlation $\rho$ follows a certain distribution $\mathcal A$ within the range $[0,1)$. Besides, the multipliers are updated by capitalizing on the sub-gradient method as
\begin{equation}\label{eqn:optstepa}
{\lambda _{s + 1}} = {\left[ {{\lambda _s} + {\theta _{{\lambda },s}}({{\rm E}_{\rho  \sim \mathcal A}}\left\{\log ({P_{out,K}})\right\} - \log (\varepsilon))} \right]_ + },
\end{equation}

\begin{equation}\label{eqn:optstepb}
{\upsilon _{s + 1}} = {\left[ {{\upsilon _s} + {\theta _{{\upsilon },s}}({{\rm E}_{\rho  \sim \mathcal A}}\left\{p_{\rm avg}\right\} - \bar p)} \right]_ + },
\end{equation}
where ${\theta _{{\lambda },s}}$ and ${\theta _{{\upsilon },s}}$ correspond to the step sizes, and $[x]_+ = \max\{0,x\}$. The pseudocode of GCN-based power allocation scheme is shown in Algorithm \ref{alg:algor1}.

\renewcommand{\algorithmicrequire}{\textbf{Input:}}  
\renewcommand{\algorithmicensure}{\textbf{Output:}} 

\begin{algorithm}[h]
  \caption{GCN-Based Power Allocation Algorithm} 
  \label{alg:algor1}
  \begin{algorithmic}[1]
    \Require
        Initial values $\bf W$, $\lambda$, $\upsilon$, ${\bf V}^{(0)}$
    \Ensure
        The power allocation policy ${\bf V}^{(L)}$;
    \For {epoch $s=1,2,\cdots$}
        \State Obtain  power allocation policy from a mini-batch.
        \State Compute the policy gradient of ${\cal L}({{\bf{W}}_s},{\lambda _s},{\upsilon _s})$.

        \State Update the primal variable $W_{s}$ [cf. \eqref{eqn:optweight}]:
        \Statex \hspace{0.68cm}${{\bf{W}}_{s + 1}} = {{\bf{W}}_s} - {\theta _{{\bf{W}},s}} {\nabla _{\bf{W}}}{{\rm E}_{\rho  \sim \mathcal A}}\left\{\mathcal L({{\bf{W}}_s},{\lambda_s} ,{\upsilon_s})\right\}$.

        \State Update the dual variable $\lambda _{s}$ and $\upsilon _{s}$ [cf. \eqref{eqn:optstepa}-\eqref{eqn:optstepb}]:

        \Statex \hspace{0.68cm}${\lambda _{s + 1}} = {\left[ {{\lambda _s} + {\theta _{{\lambda },s}}({{\rm E}_{\rho  \sim \mathcal A}}\left\{\log ({P_{out,K}})\right\} - \log (\varepsilon))} \right]_ + }$,
        \Statex \hspace{0.68cm}${\upsilon _{s + 1}} = {\left[ {{\upsilon _s} + {\theta _{{\upsilon },s}}({{\rm E}_{\rho  \sim \mathcal A}}\left\{p_{\rm avg}\right\} - \bar p)} \right]_ + }$.
    \EndFor
  \end{algorithmic}
\end{algorithm}

\section{Numerical Experiments}\label{sec:EXPERIMENTS}
Numerical experiments are conducted for verification in this section. 
For illustration, we assume that ${\xi _1} =  \cdots =  {\xi _K} = 1$, $\delta  = 1$, $K=3$, $R = 2$ bps/Hz, $N_b=10^6$~bits, $B=10$~MHz, and $\varepsilon = 10^{-2}$. With regard to the neural network structure, a $5$-layer GCN with intermediate feature dimensions 16, 32, 16 and 2 is implemented. The activation functions $\sigma ( \cdot )$ in the intermediate layers use ``\emph{ReLU}'', while the last layer applies ``\emph{Linear}''. A dataset with 1000 samples is used in the training stage, the total number of training epochs is set to $500$, and the learning rates of ${\theta _{{\bf{W}},s}}$, ${\theta _{\lambda,s}}$, and ${\theta _{\upsilon,s}}$ are assumed to be $5 \times {10^{ - 4}}$, $10^{-3}$ and $5 \times {10^{ - 5}}$, respectively. The GCN parameters $\bf W$ are updated by using the adaptive moment estimation (Adam) optimizer. Besides, the expectations in \eqref{eqn:optweight} - \eqref{eqn:optstepb} are taken over the sampled mini-batch of size 50, and $\rho $ is randomly generated from a uniform distribution within the interval $\left[ {0,1} \right)$. 

\begin{figure}[htbp]
    \centering
    \includegraphics[width=8cm]{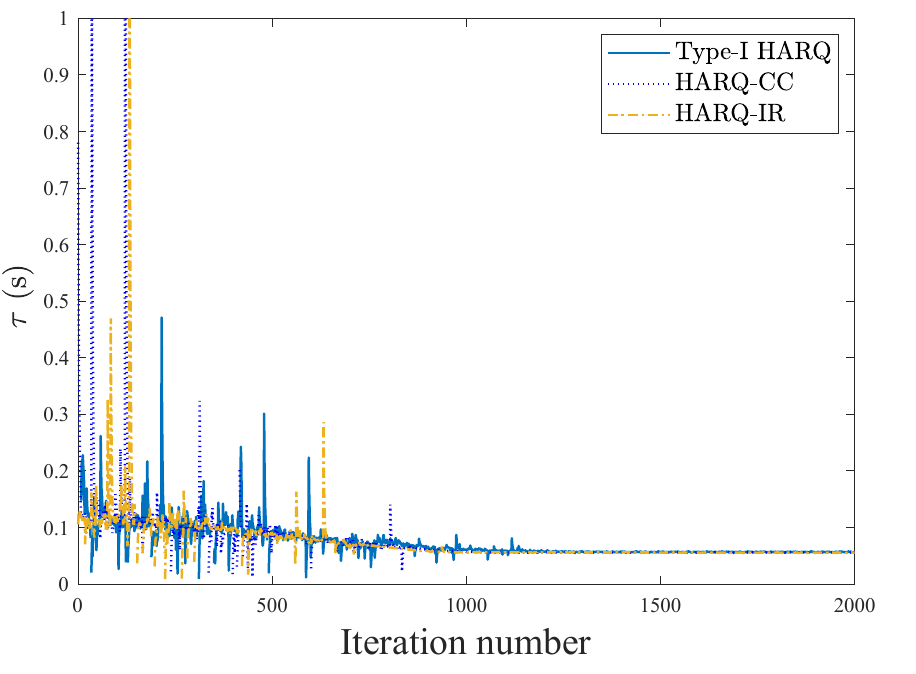}
    \caption{The convergence analysis of the primal-dual learning algorithm with respect to the number of iterations.}
    \label{FIG7} 
\end{figure}

In Fig. \ref{FIG7}, the convergence of the primal-dual learning algorithm for three HARQ schemes is investigated by setting $\bar p = 15$~dBW. Clearly from Fig. \ref{FIG7}, the proposed algorithm can converge within $1200$ iterations, which justifies the effectiveness of the GCN-based power allocation strategy. If the maximum power constraint $\bar p$ is sufficiently large (e.g., 15~dBW), the LTAT converges to the predefined transmission rate $R=2$~bps/Hz. Hence, it is observed from Fig. \ref{FIG7} that the latency approaches to $N_b/(B\eta)=10^6/(10^7\times 2)=0.05$~s with the increase of the number of iterations.


\begin{figure}[htbp]
    \centering
    \includegraphics[width=8cm]{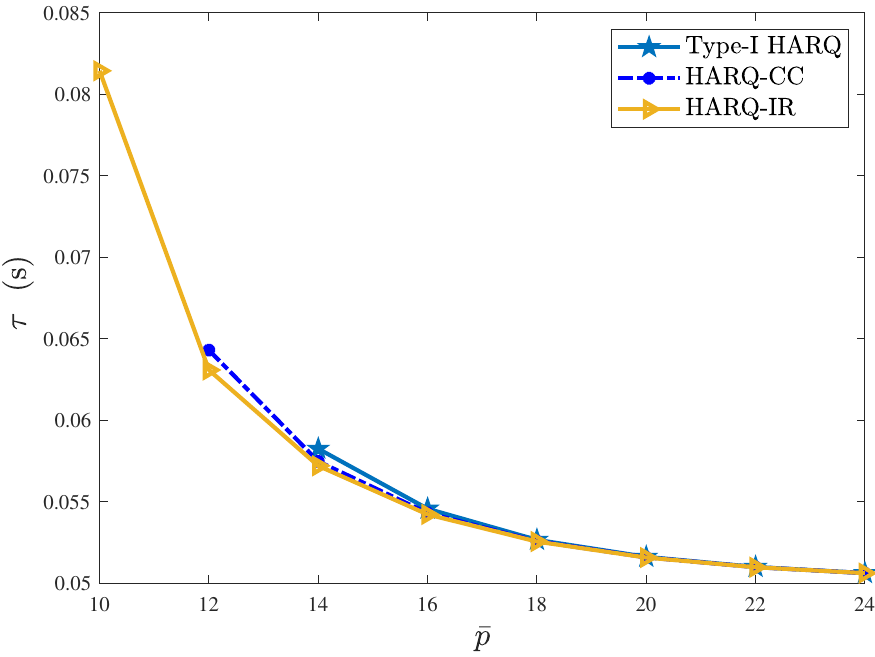}
    \caption{The comparison between the latency of different HARQ
schemes.}
    \label{FIG4} 
\end{figure}

\begin{figure}[htbp]
    \centering
    \includegraphics[width=8cm]{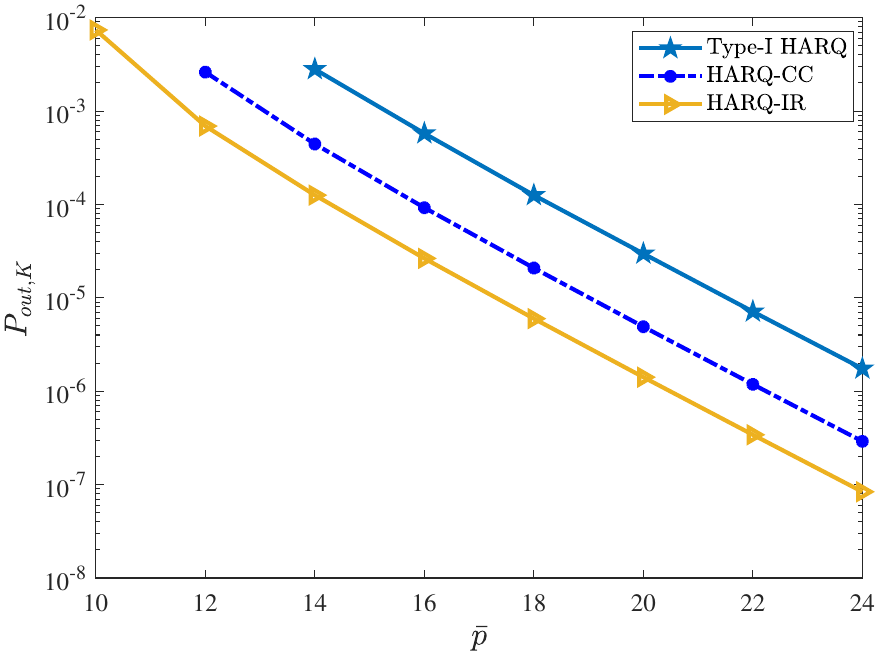}
    \caption{The comparison between the outage probabilities of different
HARQ schemes.}
    \label{FIG3} 
\end{figure}

In Figs. \ref{FIG4} and \ref{FIG3}, the minimal latency and the corresponding outage probability are plotted against the total average transmit power $\bar p$, respectively, where $\rho=0.5 $ is considered. It is not beyond our expectation from both figures that HARQ-IR performs the best, followed by HARQ-CC, and the worst is Type-I HARQ. Moreover, it can be seen from Fig. \ref{FIG4} that a significant latency reduction can be achieved by HARQ-IR under a small $\bar p$ when compared to the other two HARQ schemes. In addition, it can be observed from Figs. \ref{FIG4} and \ref{FIG3} that there is no feasible solution for Type-I HARQ and HARQ-CC under a sufficiently low transmit power. However, under a large power constraint, e.g., $\bar p>16$~dBW, the latency curves of three HARQ schemes almost coincide with each other. Hence, the superior performance of HARQ-IR in terms of the low latency is weakened as $\bar p$ increases. 
Nevertheless, it can be seen from Fig. \ref{FIG3} that HARQ-IR still has a notable outage reduction compared to the other two schemes. Moreover, as $\bar p$ increases, the latency is lower bounded by 0.05~s, which has been illustrated in Fig. \ref{FIG7}. Whereas, the corresponding outage probabilities of three HARQ schemes continuously decline with $\bar p$.

\begin{figure}[htbp]
    \centering
    \includegraphics[width=8cm]{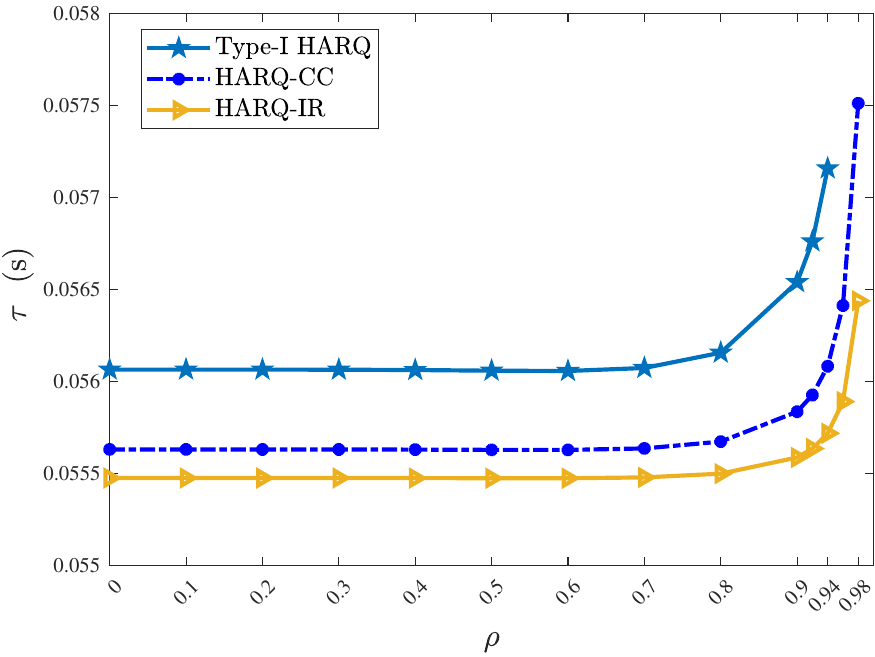}
    \caption{Effect of the time correlation on the latency.}
    \label{FIG5} 
\end{figure}

\begin{figure}[htbp]
    \centering
    \includegraphics[width=8cm]{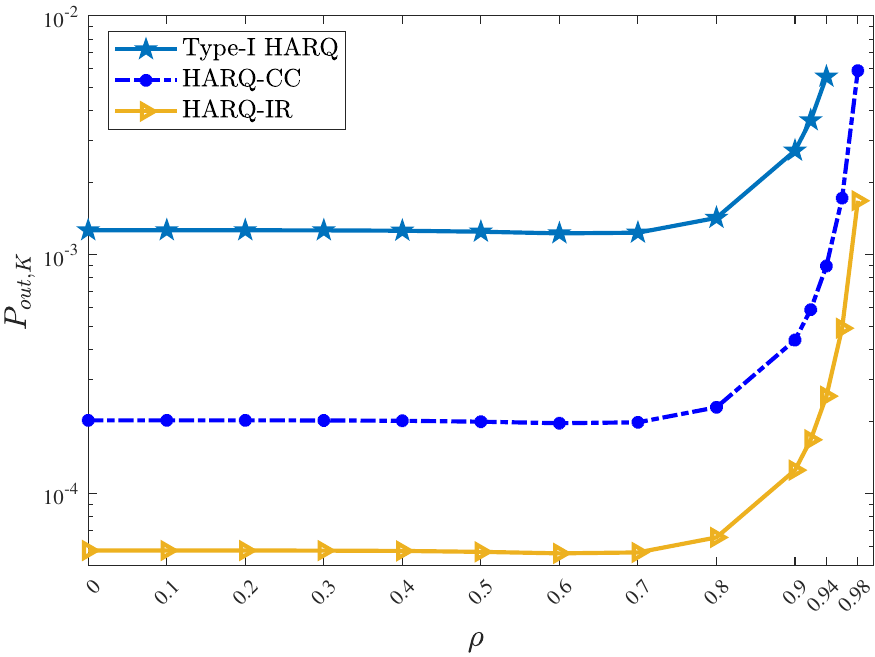}
    \caption{Effect of the time correlation on the outage probability.}
    \label{FIG6} 
\end{figure}

As shown in Figs. \ref{FIG5} and \ref{FIG6}, the effects of the time correlation on the latency and the outage probability are respectively examined by fixing ${\bar p}=15$~dBW. It is consistent with the observations in\cite{7448651,7833558,7959548} that the time correlation has a negative impact on the latency and outage performance. For example, as the time correlation increases from 0 to 0.98, the latency of HARQ-IR increases from 0.0554s to 0.0564s, and the corresponding outage probability of HARQ-IR decreases from $5.76 \times {10^{ - 5}}$
 to $1.68 \times {10^{ - 3}}$. Moreover, when the HARQ channels undergo a slightly correlated fading, i.e., $\rho<0.5$, the impact of the time correlation on the latency and the reliability can be disregarded. To sum up, HARQ-IR has the superior performance to offer low latency while ensuring high reliability, albeit at the price of extra coding complexity.

\section{Conclusion}\label{sec:CONC}
This paper studied the power-constrained HARQ schemes for realizing URLLC. More specifically, the transmission latency of HARQ schemes was minimized while guaranteeing the high reliability and limited power consumption. To render the optimization tractable, the asymptotic outage expression was used. By considering the intricate relationship between different HARQ rounds, the GCN was invoked to enable the latency minimization problem owing to its capability of tackling the graph data. The primal-dual learning method was then leveraged to train GCN parameters. Finally, the numerical experiments were performed to corroborate the validity of the proposed power-constrained HARQ schemes and compare the latency and reliability performance among three HARQ schemes. 


\bibliographystyle{IEEEtran}
\bibliography{reference}
\end{document}